\documentclass{revtex4}
\usepackage{color}
\usepackage{times}
\usepackage{graphicx}
\usepackage{fancyhdr}
\usepackage{float}
\usepackage{multirow}
\usepackage{amssymb}
\newcommand{\GW}{gravitational wave}
\newcommand{\tunit}{$\;\textrm{Myr}^{-1}\,\textrm{L}_{10}^{-1}$}

\begin{document}

\title{Chance Estimations for Detecting Gravitational Waves with LIGO/Virgo\\ Associated with Gamma Ray Bursts}
\author{Alexander Dietz\\LAPP, Universit\'{e} de Savoie, CNRS/IN2P3,\\
Chemin de Bellevue, BP 110,\\ 74941 Annecy-le-Vieux Cedex, France}

\begin{abstract} 
Short Gamma Ray Bursts (SGRB) are believed to originate from the merger of two compact objects. If this scenario is correct, SGRB will be accompanied by the emission of strong gravitational waves, detectable by current or planned GW detectors, such as LIGO and Virgo. No detection of a gravitational wave has been made up to date.
In this paper I will use a set of SGRB with observed redshifts to fit a model describing the cumulative number of SGRB as a function of redshift, to determine the rate of such merger events in the nearby universe. These estimations will be used to make probability statements about detecting a gravitational wave associated with a short gamma ray burst during the latest science run of LIGO/Virgo.
Chance estimations for the enhanced and advanced detectors will also be made, and a comparison between the rates deduced from this work will be compared to the existing literature.
\end{abstract}

\maketitle

\section{Introduction}

Gamma Ray Bursts (GRB) are intensive bursts of high-energy gamma rays, distributed uniformly over the sky, lasting milliseconds to hundreds of seconds. Several thousands of bursts has been discovered to date, with the very prominent feature of a bimodal distribution of the durations of the bursts, with a minimum around 2 seconds \cite{Kouveliotou:1993,Horvath:2002}. 
Bursts with a duration shorter than 2 seconds are called short GRB's, and bursts lasting longer than 2 seconds are labelled long GRB. Long GRB's have been associated with star-forming galaxies \cite{Jakobsson:2006,Watson:2006,Kawai:2006} and are believed to be created by stellar core-collapse. In fact, several long GRB's have been associated with observed supernovae \cite{Campana:2006,Malesani:2004,Hjorth:2003,Fruchter:2006,Woosley:2006}.

The origin of shorts GRB's, on the other hand, had been a mystery for a long time, although some time ago it has been proposed that they arise from the merger of two compact objects,  like a binary neutron star (BNS) or a neutron star-black hole binary (BHNS)\cite{eichler:1989, narayan:1992}. This picture for short GRB's is mostly accepted today, especially strengthened by recent observations of the SWIFT satellite \cite{SWIFT,swift1,swift2}:

First, while some of SGRB occur within starforming galaxies (like GRB050709), other SGRB's are being found in the outskirts of elliptical galaxies, without ongoing star formation (e.g. GRB050509B \cite{gehrels:2005}). 
This can be explained by the assumption that the merger gets a high kick velocity when one of the components undergoes a supernova explosion. Then, due to the long time scale of ~10-1000 Myr before the two components will merge, the system has enough time to leave the host galaxy and create a (S)GRB outside, which is expected to happen for a substantial fraction of mergers  \cite{belczynski06}. Also, given the long time scale, the galaxies themselves have enough time to evolve to a late-type galaxy. 

Another support for the merger model comes from SWIFT observations of weak afterglows for some short GRB's \cite{fox:2005,hjorth:2005,Stratta:2007}, which are much weaker than that of long GRB's. The weaker afterglow can be explained by the lower energy emitted in a SGRB, but also by an environment with a much lower density surrounding the GRB.  This is consistent with both the merger hypothesis \cite{piro:2005,lee:2005} and the assumption that afterglow radiation is created by external shocks.
This evidence favors the model that SGRB's originate from the merger of two massive objects, taking place in the outskirts of evolved galaxies, with the mechanism for creating the electromagnetically radiation the same as assumed for long GRB's (internal shock model) \cite{meszaros:2006,nakar:2007}. Therefore, the time-scale between any outgoing gravitational wave and the onset of the electromagnetic radiation should not exceed some milliseconds. 

Besides the merger scenario it is assumed that some of the short GRB's are caused by soft gamma repeaters (SGR), fast rotating magnetically-powered neutron stars, creating 'star quakes' in the crust from time to time and generating bursts of gamma radiation \cite{Mereghetti:2008, woods:2004}. It has been estimated that up to $\sim$25~\% of all SGRB's are caused by SGR's \cite{Tanvir:2005,Levan:2008}.
If a reasonable large fraction of SGRB's is indeed created by the merger of two compact objects, they represent putative sources of gravitational waves and could produce a measurable signal in current or planned gravitational wave (GW) detectors, such as LIGO and Virgo. 

The LIGO detectors, described in detail in \cite{LIGOS1instpaper,Barish:1999}, consist of three kilometer-length, orthogonal interferometers at two sites. One detector is located at Hanford, WA (US) and the other at Livingston, LA (US). Both detectors contain a 4~km long interferometer, while the Hanford detector additionally contain a 2 km interferometer housed in the same vacuum tube. 
The Virgo detector  \cite{0264-9381-23-19-S01}  is located near Pisa in Italy and consists of a 3 km long interferometer. 
All detectors work at or close to their design sensitivity and the LIGO detectors recently finished a 2-year data-taking run (S5) from November 2005 to October 2007. Virgo joined this data-taking effort in May 2007. 
Several results on searches for a merger signal have been published on data taken in earlier science runs \cite{LIGO03,LIGOS2iul,LIGOS2macho,LIGOS2bbh,DAbbott:2007c} and analysis is finishing for the recent S5 run. An analysis using S5 data for a search of GW associated with GRB 070201 has been performed \cite{grb070201,DietzProcGRB}, but so far, no GW has been detected. See \cite{camp:2004} for an overview on gravitational wave astronomy and the status of the facilities.

This paper will first describe the model fitted to the observed data, before the data and the fit results are presented. A comparison with rate estimations from pulsar observations and population synthesis follows, before chance estimations for the detection of GW's in current and planned detectors will be given.

\section{Description of the model}

The aim of this section is to describe a most general model for the rate of astronomical objects, as a function of their cosmological redshifts. These models follow descriptions by Chapman \cite{chapman:2007,chapman:2008} and Guetta \cite{guetta:2005} and will be used to model the distribution of short GRB's in the following section. This section does not contain any new outcomes, but for clarity and for completeness every piece of this model is shown in details.  

This model, stating the number of objects with a redshift smaller than some redshift $z_*$, is:

\begin{equation}
 N( z_*) \propto \int_0^{z_*} dz\;\frac{R(z)}{1+z} \; \frac{dV(z)}{dz} \; \int_{L_{min}(P_{lim},z)}^{L_{max}} \Phi(L) dL\;. \label{eq:generalFitfunction}
\end{equation}

In this equation $N(z_*)$ is the number of SGRB with a redshift smaller than $z_*$, $R(z)$ is the rate-function (in units per volume) at a redshift $z$, $\Phi(L)$ is the luminosity function and $dV(z)/dz$ is the volume of a co-moving shell at redshift $z$.
The integration over the luminosity function is taken out between a minimum luminosity $L_{min}(z)$, determined by the distance of the source and the satellite threshold,  and an constant upper bound $L_{max}$ for computational reasons (see Section \ref{sec:lumFunction} for details).

The rate function $R(z)$ describes the change of the intrinsic rate of objects as a function of redshift $z$ using several approaches: a function that follows the star-formation rate, and two functions following a delayed star-formation rate. They will be described in Section \ref{sec:rateFunction} in more detail. 

The luminosity function $\Phi(L)$ describes the distribution of sources as a function of their luminosities; this could follow a single power-function, a Schechter function or a log-normal distribution. These functions are described in more detail in Section \ref{sec:lumFunction}, with a detailed derivation of the norming of these functions in Appendix \ref{app:norming}. These functions have one to three free parameters, which are the ones being fitted.

\begin{figure}[t]
\includegraphics[width=4in,angle=0]{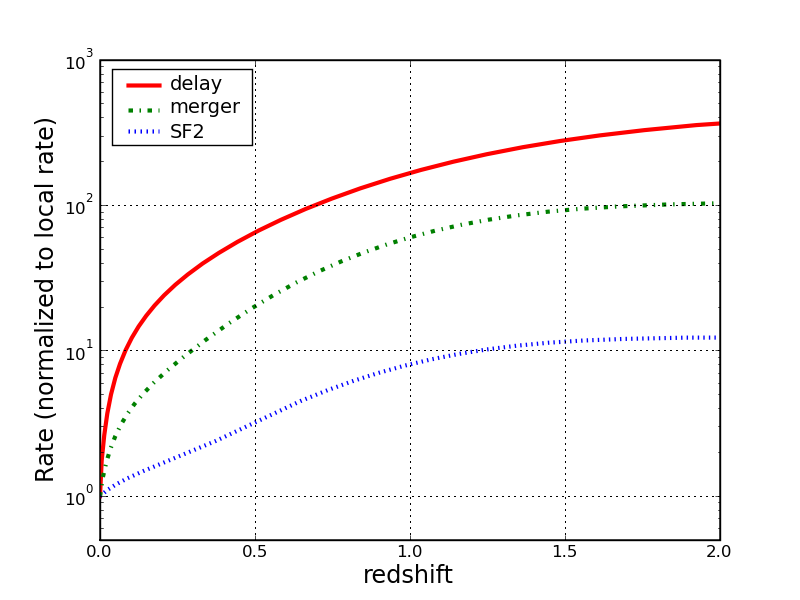}
\caption{Illustration of the different rate functions used for the models which are fitted to the observed data (see \cite{porciani:2001,chapman:2008} for details). Each model is scaled to unity at zero redshift.}
\label{fig:compareRate}
\end{figure}

\subsection{The rate function}
\label{sec:rateFunction}

This section describes the rate functions which are used to fit the model to the data given in eq.  (\ref{eq:generalFitfunction}). I will follow the same notation and enumeration as used in \cite{guetta:2005}: 

\begin{enumerate}

\item A rate that follows the star formation rate (model ``SF2`` as described in \cite{porciani:2001,guetta:2005}):
\begin{equation}
 R(z)\equiv R_{SF2}(z)= R_{s,0}\frac{23 \exp{(3.4\,z)}}{\exp{(3.4\,z)}+22}
\end{equation}

\item A rate following the merger rate of two compact objects, as derived in \cite{guetta:2005} from six observed double neutron stars \cite{champion:2004}. This rate is following a time-delay distribution ($\propto 1/\tau$):

\begin{equation}
 R(z)=R_{M,0} \int_0^{t(z)} d\tau \,R_{SF2}\left(z(t-\tau)\right)/\tau 
\end{equation}

\item A similar rate that follows SF2 with a \textit{constant} time-delay distribution :

\begin{equation}
 R(z)  = R_{D,0} \int_0^{t(z)}  d\tau\; R_{SF2}\left(z(t-\tau)\right)  \;.
\end{equation}

\end{enumerate}

In these equations $t(z)$ is the look-back time for a redshift $z$ and $z(t)$ the inverse of that. 
Fig. \ref{fig:compareRate} compares these rate functions up to a redshift of 2 (scaled to be unity for zero redshift).

\begin{figure}[b]
\includegraphics[width=4in,angle=0]{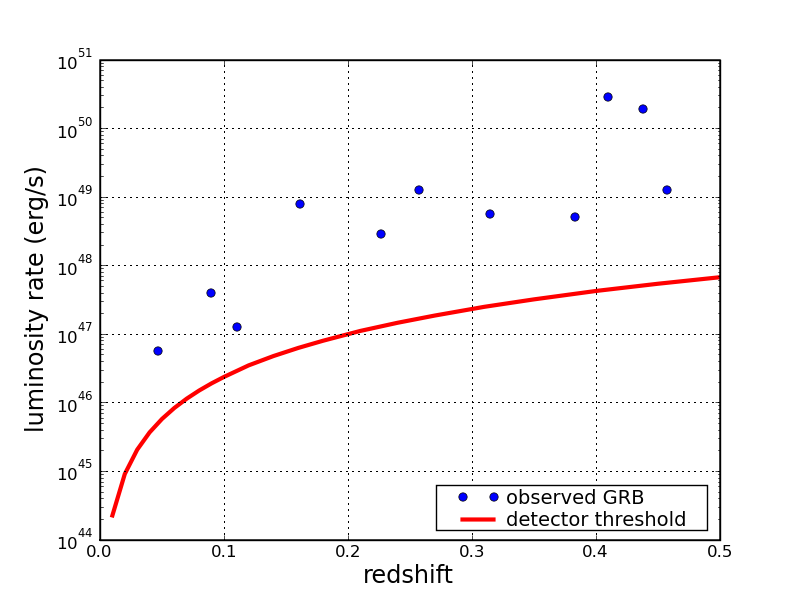}
\caption{ Plot of the approximate luminosity rate against the measured redshift for all the short GRB from Table \ref{tab:set1} as blue circles. Also shown (solid red line) is the detector threshold, assuming a required minimum flux of $10^{-8} ergs/cm^2$ to produce a trigger (see Figure 12 in \cite{sakamoto:2008}). The area \textit{below} the solid curve is not accessible by the instruments, and consistently all the GRB's lie above this line.
The most left dot in this plot is GRB 080121, located at a very small redshift of 0.046 with a consistent luminosity \cite{grb080121}. }
\label{fig:grbLum}
\end{figure}

\subsection{The luminosity function}
\label{sec:lumFunction}

The luminosity function describes the distribution of luminosities of the sources, for which often a power-law function or a schechter function is used in astrophysical context.  
These functions define the percentage $\eta$ of the sources detectable by detectors such as HETE-II and SWIFT:

\begin{equation}
\eta = \frac{1}{\Phi_0} \int^{L_{max}}_{L_{min}(P_{lim},z)} \Phi(L) \;dL
\label{eq:eta}
\end{equation}

This function describes the fraction of the sources which can be seen by a detector, given the luminosity function $\Phi(L)\equiv dN/dL$ and a norming constant $\Phi_0$. The lower limit of this integral depends on the detector threshold, which is $\sim 1\;\textrm{photon}/\textrm{cm}^2/\textrm{sec}$ for both HETE-II or SWIFT \cite{guetta:2005,sakamoto:2008}, or roughly  $P_{lim}\approx 10^{-8} ergs/cm^2$ (see Figure 12 in \cite{sakamoto:2008}). The relationship for the minimum luminosity is simply

\begin{equation}
L_{min}(z) = \frac{4\pi}{1+z} D_{lum}^2(z)\, P_{lim}
\end{equation}

Given the redshift of the source the corresponding minimum luminosity can be calculated, which acts as the lower integration limit in eqs. (\ref{eq:generalFitfunction}) and (\ref{eq:eta}).
The upper bound for this integration, $L_{max}$, is set to $10^{55}$ ergs, which has been shown to be a reasonable value - using a larger upper limit has little effect on the outcome of the fits.

The following list summarizes the different luminosity functions that are considered in the fitting model  (\ref{eq:generalFitfunction}):

\begin{enumerate}
 \item A single power law distribution, which is often used to describe the pdf of luminosity in astrophysics (two parameters: $\Phi_0$ and $\alpha$):

\begin{equation}
 \Phi(L)=\Phi_0\, \left( \frac{L}{L_0}\right)^{-\alpha}
\end{equation}

 \item A broken power law distribution, describing e.g. two underlying populations in the luminosity \cite{guetta:2005} (four parameters: $\Phi_0$, $L_0$, $\alpha$ and $\beta$):

\begin{eqnarray}
\label{eq:broken}
 \Phi(L)= &\Phi_0\,  \left( \frac{L}{L_0}\right)^{-\alpha} &\quad \textrm{for}\; L<L_0 \\
 \Phi(L)= &\Phi_0\,  \left( \frac{L}{L_0}\right)^{-\beta}  &\quad \textrm{for}\; L>=L_0
\end{eqnarray}

\item The Schechter distribution, as used for example in ref \cite{Andreon:2006} (three parameters: $\Phi_0$, $L_0$ and $\alpha$):
\begin{equation}
 \Phi(L)=\Phi_0\, \left( \frac{L}{L_0}\right)^{-\alpha}  \exp(-L/L_0)
\end{equation}

\item A log-normal distribution, describing a standard candle, e.g. a population with about the same luminosity (following \cite{chapman:2008}, with three parameters: $\Phi_0$, $L_0$ and $\sigma$):

\begin{equation}
 \Phi(L)=\Phi_0\, \frac{1}{L}\,\exp\left(\frac{-(\log{L}-\log{L_0})^2}{2\sigma^2}\right)
\end{equation}

\end{enumerate}

The constant $\Phi_0$ is always chosen so that the integral between the lower limit $L_\vee$ and the upper limit $L_\wedge$ for the luminosity is unity: $\int_{L_\vee}^{L_\wedge} \Phi(L) dL=1$. 
The used integration limits are  $L_\vee=10^{43}\;$ergs and $L_\wedge \equiv L_{max}=10^{55}\;$ergs. These integration limits have been chosen because the derived luminosities for all used SGRB's given in table \ref{tab:set1} range between $10^{47}$ ergs and $10^{50}$ ergs, well within the integration limits. It also has been shown that the results of the fit do not change significantly when widening the integration limits in either direction.
The derivation of the constants $\Phi_0$ can be found in Appendix \ref{app:norming}.

\section{Used data and fit results}
\label{sec:fitresults}

\begin{table*}[b]
\begin{tabular}{|l|l|l|l|l|}
\hline
 \textbf{GRB}    & \textbf{redshift}  & \textbf{duration [s]} & \textbf{fluence [$10^{-7} \frac{\textrm{erg}}{\textrm{cm}^2}$]} & \textbf{luminosity [$10^{49}$ erg]} \\ \hline
050509B\cite{nakar:2007}      & 0.226   & 0.040 & 0.095 & 0.012\\ \hline 
050709\cite{nakar:2007}       & 0.1606  & 0.22  & 10.0   & 0.175  \\ \hline
050724\cite{nakar:2007}       & 0.257   & 1.31  & 9.9  & 1.67  \\ \hline
060505\cite{grb060505}             & 0.089   & 4.0 & 6.2  & 0.16 \\ \hline
061006\cite{grb061006,berger:2007} & 0.4377 & 0.42 & 14.2 & 8.27 \\ \hline
061210\cite{grb061210,berger:2007} & 0.4095 & 0.19 & 11   & 5.52 \\ \hline 
070209\cite{grb070209}             & 0.314  & 0.10 & 0.22 & 0.059 \\ \hline
070406\cite{grb070406}             & 0.11   & 0.70 & 0.36 & 0.009 \\ \hline
070724\cite{grb070724}             & 0.457  & 0.40 & 0.80 & 0.517 \\ \hline 
071227\cite{grb071227}             & 0.383  & 1.8  & 2.2  & 0.933 \\ \hline 
080121\cite{grb080121}             & 0.046  & 0.7  & 1.0  & 0.004 \\ \hline  
\end{tabular}
\caption{Set of 11 short gamma ray bursts with certain redshift measurements and $T_{90}$ duration equal or less than 4 seconds, and with a redshift smaller than 0.5, used for the fit. The data for these GRB's are taken from \cite{oshaughnessy:2007,Sakamoto:2007} and from diverse GCN circulars (see \cite{gcn}). The fluence and the luminosity of these GRB's are listed as well, but this information is not used for the fitting itself in this work. The luminosity has been calculated using the fluence and the luminosity distance to each GRB.}
\label{tab:set1}
\end{table*}

\begin{figure}[ht!]
\includegraphics[width=4.0in,angle=0]{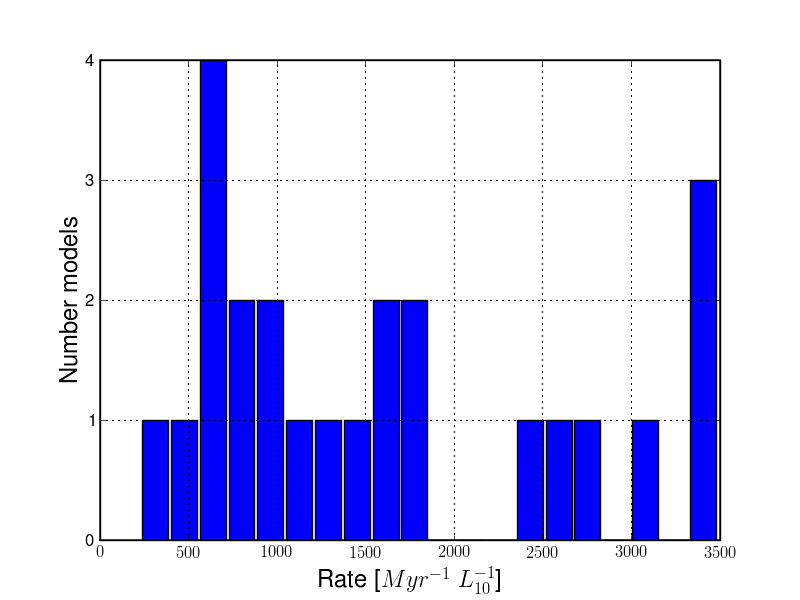}
\caption{Histogram of the distribution of rates from the 24 accepted fit results, as described in the text. The rate values for this plot are also listed in the last column of Table \ref{tab:fitresult2}.}
\label{fig:histRates}
\end{figure}

\subsection{Used data set}

The data used to perform the fit consist of 11 short GRB's \footnote{i.e. with a $T_{90}$ duration less/equal 4 seconds to account for possible SGRB's with a somewhat longer duration} with well-defined redshift measurement up to z=0.5, detected by various satellite missions between May 2005 and September 2008. Some properties of these GRB's are given in Table \ref{tab:set1}. 
This table include fluences and luminosities for the GRB's as well, which are not used in the fitting procedure, because these values differ much between the instruments and the energy ranges; they are only used for a consistency check (see below). The cumulative number of the GRB's in Table \ref{tab:set1} is used to fit the model given in eq. (\ref{eq:generalFitfunction}).

Figure \ref{fig:grbLum} shows the distribution of the luminosity rates (the luminosity divided by the $T_{90}$ time) of the SGRB's and the solid red line corresponds to the approximate threshold of the satellites (using a ballpark threshold of $10^{-8} ergs/cm^2/s$). As expected (and for a sanity check), all GRB's lie above the threshold line, including GRB 080121 (the most left dot in this plot)  with a redshift of only 0.046. This particular short GRB has a very low luminosity, several orders of magnitude less than typical short-hard bursts \cite{grb080121}, but is consistent with its distance.

\subsection{Results of the fit}

The fit of the model was performed by a least-squares fitting function from the scipy module of python \cite{scipy}, which uses a modified version of the Levenberg-Marquardt algorithm to minimize a given function, which is the difference between the observed data and the fit (eq. (\ref{eq:generalFitfunction})). This fit depends on some parameters: one overall normalization factor and one to three parameters used in the luminosity function (see Sec. \ref{sec:lumFunction}). Given these fit-parameters the $\chi^2$-value of the fit is calculated, as well as the Kolmogorov-Smirnov (KS) probability, computed by taking the largest difference between the cumulative histogram of measured and estimated datapoints. From this difference a probability $\alpha$ is derived, at which the null hypothesis (i.e. samples are drawn from the same distribution) is \textit{not} rejected (see e.g. \cite{numericalrecipes} for details).  

Besides performing the fit using only \textit{one} luminosity function, fits were performed including \textit{two} different luminosity functions in the case of more than one underlying population (see e.g. \cite{chapman:2008} for a similar approach).
This might be the case, because of the contribution of SGR's to short GRB's, which do not originate from a merger.

To take into account only reasonable models, fits with a KS probability less than 95\% or with a $\chi^2$ value per d.o.f. of more than 5 have been rejected. 
Because an estimated fraction of $\sim$25\% of all SGRB's are caused by SGR's \cite{Tanvir:2005,Levan:2008}, the major population of SGRB's is likely to come from  the merger of two compact objects. Therefore the model with the \textit{larger} rate prediction is chosen for any two-luminosity-model fit. 
Table \ref{tab:fitresult2}  summarizes the final sample of models, including their fitted parameters and the goodness-of-fit values. 
It is not surprising that many models with the SF2 rate-model fit the data well, because it has been shown that the distribution of SGRB's should follow the SF2 rate pretty closely \cite{belczynski:2002,belczynski06}.

To obtain a \textit{rate} estimation for a model, each model (i.e. equation (\ref{eq:generalFitfunction})) is integrated up to a distance of 100~Mpc (corresponding to a redshift of  0.02258), and rescaled by the number 11 of GRB's used in the fit. This number corresponds to the probability that a given GRB lies within 100~Mpc. 
This number must be further multiplied by a factor 15/28800$\times$100. 15 is the approximate number of short GRB's occurring during one year, 28$\;$800 is the approximate number of $L_{10}$ within 100~Mpc \footnote{$L_{10}$ is $10^{10}$ times the blue solar luminosity. Our galaxy contains about $1.7L_{10}$ \cite{Kalogera:2000dz}.}, and $f_b^{-1}=100$ is the approximate beaming factor of a SGRB  \cite{nakar:2007,guetta:2008}, to take into account all arbitrarily oriented mergers as well. The values of the rates calculated that way are given in the last column of Table  \ref{tab:fitresult2}. 

Figure \ref{fig:histRates} shows the distribution of the rates of these 24 models, spreading over a large range of values. This is not a surprise when considering that the fit itself is performed up to a redshift of 0.5, but the rate is estimated within a redshift range of only $z\simeq0.03$. 
It also should be noted that the outcome of the fits is very sensitive on the given input data; using only 10 out of the 11 GRB's to fit a model can change the rate estimation by an order of magnitude. 
Plots of all accepted models and the data are shown in Figs. \ref{fig:imageFitsSFR1}-\ref{fig:imageFitsMerger} in the Appendix.

\begin{table*}
 \begin{tabular}{|l||l|l|l||l|l|l||l|l||l|}
 \hline
\textbf{Model} & \multicolumn{3}{|c||}{\textbf{first model}} & \multicolumn{3}{|c||}{\textbf{second model}} & \textbf{KS} & \textbf{Chi} & \textbf{rate} \\ \hline \hline
                       & $L_0$ & $\alpha,\;\sigma$ & $\beta$ & $L_{break}$ & $\alpha,\;\sigma$ & $\beta$ &&& \tunit  \\ \hline \hline
SF2 power&--- & 2.0 & --- & --- & --- & --- & 98.7 & 0.15 &2530 \\ \hline
SF2 power/power&--- & 2.0 & --- & --- & 2.2 & --- & 99.1 & 0.18 &1710 \\ \hline
SF2 power/schechter&--- & 2.1 & --- & 49.1 & -16.9 & --- & 99.8 & 0.16 &3480 \\ \hline
\textbf{SF2 power/broken}&--- & 2.1 & --- & 48.0 & 1.1 & 0.8 & 99.8 & 0.19 &3490 \\ \hline
SF2 schechter/schechter&47.6 & 0.2 & --- & 50.3 & -0.2 & --- & 96.4 & 0.12 &670 \\ \hline
SF2 schechter/lognormal&48.7 & -8.2 & --- & 47.2 & 1.8 & --- & 97.2 & 0.10 &1070 \\ \hline
SF2 schechter/broken&48.5 & -24.4 & --- & 46.8 & -0.5 & 2.4 & 98.6 & 0.16 &570 \\ \hline
SF2 lognormal&46.8 & 14.7 & --- & --- & --- & --- & 98.6 & 0.17 &2370 \\ \hline
SF2 lognormal/lognormal&47.2 & 1.6 & --- & 50.6 & 0.4 & --- & 96.8 & 0.10 &890 \\ \hline
SF2 lognormal/broken&47.2 & 1.6 & --- & 48.3 & 1.1 & 0.5 & 96.8 & 0.13 &890 \\ \hline
SF2 broken&45.8 & -4.2 & 2.0 & --- & --- & --- & 98.0 & 0.20 &1640 \\ \hline
SF2 broken/broken&47.2 & 0.7 & 2.7 & 48.0 & 1.3 & 0.9 & 97.7 & 0.23 &430 \\ \hline
Merger power&--- & 2.2 & --- & --- & --- & --- & 99.1 & 0.14 &3160 \\ \hline
Merger power/power&--- & 2.3 & --- & --- & 2.0 & --- & 99.7 & 0.18 &3370 \\ \hline
Merger power/broken&--- & 2.1 & --- & 47.5 & -1.0 & 8.5 & 99.7 & 0.20 &1560 \\ \hline
Merger schechter/schechter&47.6 & 0.4 & --- & 50.5 & 0.2 & --- & 96.2 & 0.13 &590 \\ \hline
\textbf{Merger schechter/broken}&49.4 & -5.0 & --- & 46.9 & -0.9 & 2.7 & 96.6 & 0.19 &240 \\ \hline
Merger lognormal/broken&46.9 & -1.6 & --- & 46.0 & -0.2 & 1.4 & 98.1 & 0.14 &890 \\ \hline
Merger broken&46.0 & -13.0 & 2.3 & --- & --- & --- & 97.7 & 0.21 &1230 \\ \hline
Delay power&--- & 2.3 & --- & --- & --- & --- & 95.2 & 0.18 &2740 \\ \hline
Delay power/power&--- & 2.6 & --- & --- & 2.3 & --- & 96.8 & 0.22 &1770 \\ \hline
Delay power/broken&--- & 2.2 & --- & 47.4 & -0.4 & 8.6 & 97.6 & 0.27 &1450 \\ \hline
Delay schechter/schechter&47.6 & 0.5 & --- & 51.0 & 0.2 & --- & 96.7 & 0.14 &590 \\ \hline
Delay lognormal/broken&46.8 & 1.7 & --- & 48.0 & 1.3 & 0.6 & 95.5 & 0.12 &830 \\ \hline
\end{tabular}
\caption{Table containing the 24 selected fit results with a KS-probability of at least~95\% and a $\chi^2$ per d.o.f. less than~5. These 24 models are being used for the further analysis. The models in bold font have the largest and smallest rates, and the value for $L_0$ are given in units of $\textrm{log}_{10}$.}
\label{tab:fitresult2}
\end{table*}

As a result of the analysis 20\% of the models give a rate larger than 2750\tunit{}, 50\% of the models yield a rate of $\gtrsim$~1450\tunit{}, while 80\% of all models predict a rate of $\gtrsim$ 670\tunit{}. 
Since the 50\% value corresponds to a median value, this value will be used in the next section to compare with rate estimations.

\section{Comparison with other rate estimates}

In this section a rough comparison is made between the rates deduced in this paper and rates estimated elsewhere. Two cases are being distinguished: The rate of mergers of two neutron stars (BNS, Binary Neutron Star) and the rate of mergers of a Black Hole and a Neutron Star (BHNS). 
For BNS the rate is deduced from known binary pulsars in our Milky Way and are expected to be realistically at 50\tunit , although they could be as high as 500\tunit{} \cite{Kalogera:2004ntx}.
The rates predicted for BHNS are much more uncertain, and have been modeled using population synthesis. Realistic rates for BHNS lie at  2.0\tunit{}, although they could be as high as 60\tunit{} \cite{OShaughnessy:2008}. 

The models used in this work cover a rate range of 240\tunit{} to 3500\tunit, with a median value at $\sim$~1450\tunit. These values are consistently larger than the previous rate estimates. 

A similar work on using GRB's to fit the local observed population has been done in \cite{guetta:2008}, quoting two rates of 130 $\textrm{Gpc}^{-1}\textrm{yr}^{-1}$ and 400 $\textrm{Gpc}^{-1}\textrm{yr}^{-1}$, respectively, for two different models on the origin of binaries. This corresponds to a rate of 460\tunit{} and 1400\tunit;  
the latter value surprisingly close to the rate of SGRB's deduced in this work.
This might be a hint that the rates estimations in \cite{Kalogera:2004tn, Kalogera:2004nt,OShaughnessy:2008} are underestimated, but given the large uncertainties involved in the fit procedures in this paper, more statistics is needed to investigate this hypothesis.
Table \ref{tab:compareRates} summarizes the median/realistic rates deduced from several methods, papers and this work.

 \begin{table}[b!]
 \begin{tabular}{|l|l|l|}
 \hline
                       & realistic/median rate \\ 
& [\tunit] \\ \hline
this work              &  1450 \\ \hline
Ref. \cite{guetta:2008} model I &  460 \\ \hline
Ref. \cite{guetta:2008} model II &  1400 \\ \hline
BNS rate  \cite{Kalogera:2000dz} &   50 \\ \hline
BHNS rate \cite{OShaughnessy:2008} &    2.0 \\ \hline
\end{tabular}
\caption{Comparison of realistic rate estimations for BNS and BHNS with the median values of the rates deduced in this work, compared to the outcome of a similar investigation by Guetta \cite{guetta:2008}. 
The first three numbers are derived from fits to observed short GRB's, while the last two rates are deduced from binary pulsar observations and population synthesis. It is interesting to note that the rates estimated from observed redshifts (rows 1 to 3) are consistently larger than the ones estimated otherwise.}
\label{tab:compareRates}
\end{table}

 \begin{table}
 \begin{tabular}{|l|l|l|l|l|}
 \hline
  \textbf{GRB}              & \textbf{detectors}  & \textbf{range [Mpc]} & \textbf{p [\%] (BNS)}& \textbf{p [\%] (BHNS)} \\ \hline
GRB051114 & H1H2 &  20.1 & 0.02 - 0.08 & 0.12 - 0.47\\ \hline
GRB051210 & H1H2 &  21.8 & 0.02 - 0.10 & 0.15 - 0.60\\ \hline
GRB051211 & H1L1 &  33.0 & 0.10 - 0.41 & 0.51 - 2.07\\ \hline
GRB060121 & H1L1 &   5.3 & 0.00 - 0.00 & 0.00 - 0.01\\ \hline
GRB060313 & H1H2 &  21.1 & 0.02 - 0.09 & 0.13 - 0.54\\ \hline
GRB060427B & H1L1 &  56.2 & 0.49 - 1.98 & 2.49 - 9.77\\ \hline
GRB060429 & H1H2 &  33.0 & 0.08 - 0.34 & 0.51 - 2.06\\ \hline
GRB061006 & H1H2 &  21.9 & 0.02 - 0.10 & 0.15 - 0.60\\ \hline
GRB061201 & H1H2 &  30.5 & 0.07 - 0.27 & 0.40 - 1.64\\ \hline
GRB070201 & H1H2 &  15.5 & 0.01 - 0.03 & 0.05 - 0.21\\ \hline
GRB070707 & H1H2 &  31.3 & 0.07 - 0.29 & 0.43 - 1.76\\ \hline
GRB070714 & H1L1 &  17.2 & 0.01 - 0.06 & 0.07 - 0.29\\ \hline
GRB070729 & H1L1 &  52.8 & 0.41 - 1.64 & 2.07 - 8.18\\ \hline
GRB070809 & H1H2 &  10.6 & 0.00 - 0.01 & 0.02 - 0.07\\ \hline
GRB070923 & H1L1 &  19.7 & 0.02 - 0.09 & 0.11 - 0.44\\ \hline
\end{tabular}
\caption{15 short GRB's detected during the fifth LIGO Science run and VSR1 of Virgo, that have no reasonable redshift estimation, and for which data is available from at least two detectors (indicated in the second column). The third column shows the approximate detection range of the second most sensitive detector at this time, using the SNR thresholds as described and the known position of the source. 
The fourth/fifth column gives the probability range for each GRB for the BNS/BHNS case.
The total probability of a detection during S5 lies in the 20/80 range of $\sim$1.9 to $\sim$7.7\% with a median value around $\sim$ 4.1\% for the BNS case. When using the 'best' model, the total probability increases to 9.7\% (see Table \ref{tab:mcProb} for details ond for the values of the BHNS case).}
\label{tab:probAll}
\end{table}

\begin{figure}[t!]
\includegraphics[width=3.5in,angle=0]{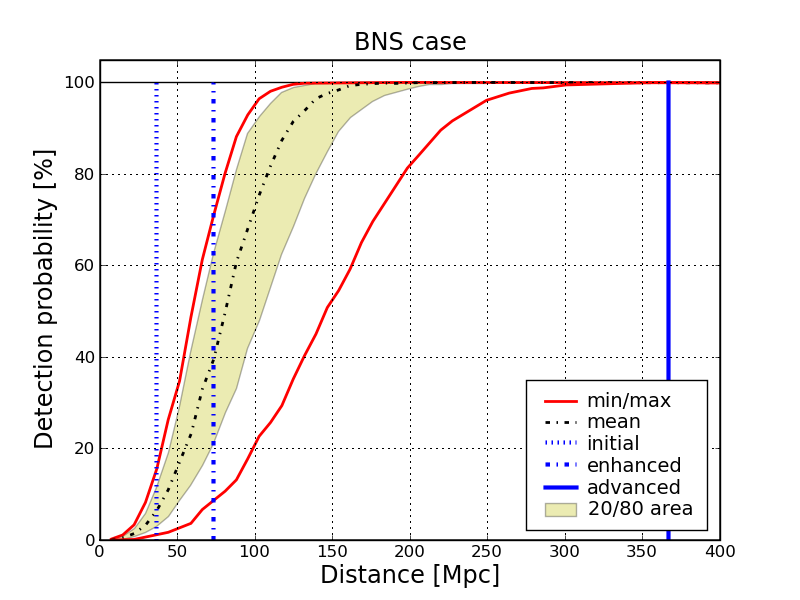}
\includegraphics[width=3.5in,angle=0]{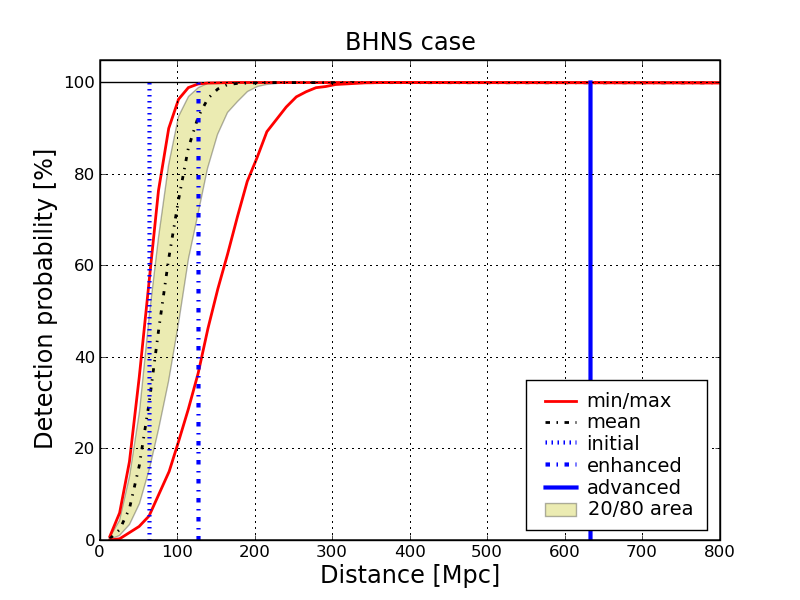}
\caption{Plot showing the detection probability as a function of detector range (left plot for the BNS scenario, right plot for the BHNS). The diversity in the detection probability in the y-direction comes from the different fits. The red curve signifies the detection probability for the 'worst' and 'best' model, while the yellow area shows the range of probability predicted by 20-80\% of the models. 
The vertical lines denote the probability areas as predicted by the models. See also Table \ref{tab:mcProb}.}
\label{fig:plotDetProb}
\end{figure}

\section{Chance estimation for LIGO and Virgo}

This section discusses the probability of a detection of gravitational waves associated with a SGRB during S5.
To calculate this probability the \textit{reachable} distance of a detector to a source must be calculated, given the known \textit{horizon} distance of a detector at a given time. 

The \textit{horizon} distance of a detector is the distance at which a merging system would create a signal in this detector with a signal-to-noise ratio of 8.0, if the system is optimally located and optimally oriented.
For the approximate calculation of the probabilities of discovering a SGRB within the analysis of S5 data (and for simplification) two cases are taken into account for the merger: a low-mass pair consisting of two 1.4  $M_\odot$ neutron stars (BNS) and a high-mass pair consisting of a 1.4 $M_\odot$ neutron star and a 10 $M_\odot$ black hole (BHNS).
Looking at the ranges of the inspiral analysis in S5 (see, e.g. \cite{MattersGravity32:2008}), the horizon distance for detecting the BNS system is about 30 Mpc for H1/L1 and 15 Mpc for H2/V1, while for the BHNS case it is 50~Mpc and 25~Mpc, respectively. 
In the following it is assumed that Virgo behaves similar to H2 with respect to the sensitivity. 

The actual search for gravitational waves of mergers from a short GRB is split up into two cases, depending on the available data, and utilizing two different thresholds. If data is available from the two most sensitive detectors H1 and L1 (L-case), the threshold is set for them to be 4.25, and in the case that one of the detectors is H2 or V1 (H-case), the threshold for the least sensitive detector is set to 3.75. 
It is very unlikely that a real gravitational wave, producing a signal just above these thresholds, will be identified as such, since it will be buried well within background noise. 
Therefore I will assume a rather conservative threshold for signal detection 50\% above the actual used threshold.

 \begin{table}[b]
 \begin{tabular}{|l|l|l|l|l|l|l|l|l|l|l|}
 \hline
             &  \multicolumn{5}{|c|}{\textbf{low mass}} & \multicolumn{5}{|c|}{\textbf{high mass}}  \\
             & worst & 80\% & 50\% & 20\% & best & worst & 80\%& 50\% & 20\%  & best\\
             &  \multicolumn{5}{|c|}{of models} & \multicolumn{5}{|c|}{of models}  \\ \hline
initial detectors (real GRB)& 0.7\% & 1.9\% &4.1\%& 7.7\% &9.7\% &  3.7\%& 9.9\% &20.2\% & 34.7\% & 41.9\% \\ \hline
initial detectors (sim GRB) &  1\%  &   3\% &7\% &  13\%  &  15\% &  7\%& 17\% & 30\%& 51\% &59.6\%\\ \hline
enhanced detectors          &  9\%  &  22\% &43\% &  65\%  &  74\% & 39\%& 75\% & 94\%& 99.3\%  & 99.8\% \\ \hline
advanced detectors          & 100\%  & 100\% & 100\%& 100\%  & 100\% &100\%& 100\%&100\%& 100\%  & 100\%\\ \hline
\end{tabular}
\caption{The outcome of a Monte Carlo simulation (except the first line) to estimate the probability to have at least one short GRB within the reachable distance of the second most sensitive detector, given 30 SGRB's within the observing time of the network. A duty cycle of 80\% for each single detector was used in the simulation.
The first line was calculated using the properties of the known GRB's during S5.
The numbers stated are the detection probabilities for the worst/best model, and the probabilities predicted by 20/50/80\% of the models. For initial LIGO the predictions are not very promising, although the best model predicts a $\sim$~60\% chance to see a detection for the BHNS case. The 100~\% values given for the advanced detectors are because of rounding effects; when doing sufficient simulations and considering the large uncertainties in the models they never reach exactly 100~\% - although they are pretty close to it. }
\label{tab:mcProb}
\end{table}

Taking this into account the distances to which a binary system can be detected for the L-case are $\sim$35 Mpc (BNS) and $\sim$65 Mpc (BHNS) and for the H-case they are $\sim$20 Mpc (BNS) and $\sim$35 Mpc (BHNS), respectively. These are, however, only rough estimates of the distance which can vary up to $\sim$ 20\% because of fluctuations in the taken data (see, e.g. the Figure showing the reach of the detectors as a function of mass in \cite{MattersGravity32:2008}).

Because a merging system is in general neither optimal located nor optimal oriented this has to be taken into account in form of the so-called antenna factors\cite{thorne.k:1987}, which depends of the sky location of the putative source and the inclination angle $\iota$ of the binary system.

This inclination angle is the angle between the axis of total momentum of the system and the direction of the line of sight from the system to earth. For a system seen from earth face-on this angle is zero, while for a system seen edge-on this angle is $90^\circ$. 
Since the axis of total momentum of the system is the only significant direction and we are able to see the electromagnetic burst, taking place in a jet within some opening angle, it is a reasonable assumption that the inclination angle is smaller than or equal to the opening angle. 
For short GRB's, the opening angle was determined to be at about 0.16 radians \cite{nakar:2007}. For the calculation of the probabilities it is assumed that the inclination angle is below $\iota=0.2$. 

Using the sky location it is possible to calculate the reachable distance for each GRB with the appropriate SNR threshold. The results of this calculation for all GRB's during S5 is shown in Table \ref{tab:probAll}.
This table also contains the probabilities that a particular GRB lie within the detectable range as predicted by 20\% (80\%) of the models. 
The total detection probability that a \GW{} will be seen in S5 data is 7.7\% for 20\% of the models, while 80\% of the models predict a probability of at least 1.9\%. The median probability lies at 4.1\% (see also Table \ref{tab:mcProb}).
The probabilities for the BHNS case lie in the 20/80 range of 9.9\% to 34.7\%, with a maximum value of 41.9\%, and a median probability of 20.2\%.

\section{Outlook for advanced detectors}

Currently LIGO and Virgo are in the process of upgrading, and a science run with the enhanced detectors will start in summer 2009. These detectors will be twice as sensitive than initial detectors \cite{whitcomb:2007} and hence can observe an eight fold larger volume in the universe. 
For advanced LIGO, supposed to start data-taking in 2014, the increase in sensitivity with respect to the initial detectors is a factor of 10 \cite{whitcomb:2007}, corresponding to an 1000-fold increase in rate. 

The chances are very good that during the runs of the enhanced and advanced detectors enough satellites are operating to deliver gamma-ray triggers. Currently three dedicated GRB missions are orbiting the earth, HETE-II\cite{hete}, SWIFT\cite{swift1,swift2} and Fermi (formerly GLAST) \cite{GLASTProcGRB}.  
While HETE-II might last some more years, the SWIFT satellite is expected to operate for the next five years, covering at least the science runs with the enhanced detectors; SWIFT may be even operational into the runs with advanced detectors. 
Also the recently started Fermi satellite, launched in June 2008, is expected to work at least 10 years \cite{GLASTlifetime}, enough time to cover data taking periods with advanced LIGO and advanced Virgo.  
Last to mention is the Interplanetary Network (IPN) \cite{IPN}, a network of spacecrafts equipped with gamma-ray detectors, such as Wind, NEAR, KONUS and Mars Odyssey. By measuring the time-delays of a trigger between these spacecraft it is possible to triangulate the approximate position of a GRB. These positions are not very precise in contrast to dedicated GRB's missions, but in most cases precise enough for an associated search of GW from this event. 

To estimate the probability of a detection with these detectors, a 2-year data taking run is simulated for each the enhanced and the advanced LIGO/Virgo detector network. 30 short GRB's have been simulated during this time, and the chance that at least two detectors will take data at the time of each GRB, assuming a duty-cycle of 80\% for a single detector. 
The randomly selected sky location defines the sensitivity to all available detectors, of which the second most sensitive is being chosen, since this detector will limit the reach of the observation. 
Then for each model (taken from table \ref{tab:fitresult2}) the chance is being calculated that at least one of the 30 SGRB's is within detectable range and at least two detectors are 'data taking'. If that is the case, a 'detection' has been made. 
This procedure is repeated 5000 times for each model to estimate the chance to have a detection with a given model. This large number of trials ensures a relatively small Monte Carlo error of about 5\%; the uncertainties in the duty-cycle and the number of SGRB per year have a much larger effect on the uncertainties. A change in the duty-cycle between 70\% to 90\% or a change in the number of SGRB's within 2 years between 25-35 yields an error of the MC results on the order of $\sim$~30\% each.

Figure \ref{fig:plotDetProb} shows the outcome of this simulation for arbitrary sensitivities or detector ranges; the sensitivities for initial, enhanced and advanced detectors are marked by vertical lines.
The spread in the detection probabilities come from the spread in the rate predictions of the different models and the yellow area shows the probability range predicted by 20-80\% of the models.

 For initial LIGO (dotted vertical blue line), the predicted detection probabilities for the BNS case range from 0\% (worst model) to 15\% (most optimistic model). The 20/80-area covers a range between 3\% and 13\%, i.e. for 20\% of the models the chance of a detection is at least $\sim$13\% and for 80\% this chance is at least 3\% (see also the second row in Table \ref{tab:mcProb}). These predictions are consistently larger than the estimates in the first row in Table \ref{tab:mcProb} using the real observed GRB's, because for most of S5 the Virgo detector was not online. 

With the enhanced detectors the changes of detecting a \GW{} associated with a GRB are much larger, and range between 22\% to 65\% (20/80-area) for the BNS case and up to 100\% for the BHNS case.  And for the advanced detectors finally, the Monte Carlo simulations predict a 100\% chance for the detection of a \GW, even for the worst model! The real chances for a detection are not exactly 100\% (which is due to finite simulations), but very close. 
Then, at last, a new window to the universe should open up. 
Table  \ref{tab:mcProb} gives a summary of the detection estimations for initial, enhanced and advanced detectors.

It has to be pointed out, as mentioned earlier, that these probabilities are based on a small data sample and therefore inherit large uncertainties. The largest uncertainties come from the small number input data itself, to which the fits are very sensitive to. Other uncertainties come from the assumed duty cycle of the detectors, the number of SGRB's detected by gamma-ray satellites during the science runs, and the SNR threshold of a putative signal.
However, taking these uncertainties into account, the probability to have a detection of a gravitational wave associated with a short GRB with the enhanced detectors is about 50\%, which is a very exciting prospects for the next science run expected to start in 2009.

\section{Summary and conclusion}

In this paper I made an attempt to estimate the local rate of short gamma ray bursts from a set of 11 short gamma ray bursts with reliable redshift measurements. I then compared them with values found in literature and used them to predict the chances of a detection of a \GW{} assuming such events are mostly created by the merger of two compact objects. 
This assumption has gathered much support during the last years by observations with satellites like Swift.

A model, describing the number of short GRB's as a function of redshift using different rate- and luminosity functions, has been used to fit the observed data. These fits, made with a modified version of the Levenberg-Marquardt algorithm, tried to minimize the difference between model and data, by altering the free parameters in the model. 
A KS test and a $\chi^2$ criterion was used to select only models with reliable good fits.    
Because these fits were made to a redshift of 0.5, but conclusions about the local rate are drawn from redshifts smaller than $\sim$~0.02, they have to be taken with large caution. 
Also, the variation of the input data has a large effect on the results and can change the results by an order of magnitude. 

Taking the median value of the fit results, which are spread over an order of magnitude, they look to be consistently larger than the predictions on the local rate made from binary pulsar observations and population synthesis. 
This is also the outcome in a similar investigation, leading to the possibility that the real local rate of mergers is larger than previously thought; but it is too early to draw firm conclusions on that.

Although the rate estimations are relatively large, the prospects to discover a \GW{} associated with a SGRB in the fifth science run in LIGO/Virgo are not promising; the probability for such a detection are well below 50\%. Unlike larger are the chances to detect a \GW{} in enhanced LIGO/Virgo: Monte Carlo simulations predicts a probability of $\sim$~50\% for the BNS case and even $\gtrsim$90\% for the BHNS case.  For advanced detectors the chance of detecting a \GW{} associated with a GRB turns out to be $\sim$~100\%, even for disadvantaged models (for both BNS and BHNS cases). This large probability has to be taken with caution again, but even taking into account all the uncertainties they are very large. 
 Such a coincident detection will not only give strong significance of the true nature of a real signal, it will also allow a much better distance estimation to cosmological sources, if the redshift of the GRB is known. 
Further observations of GRB associated \GW{}s therefore will not only increase the knowledge on stellar formation and galaxy evolution, but will have also a huge impact on the theories on large structures in the universe and on cosmology.

\acknowledgments

I would like to thank Richard O'Shaughnessy for useful discussions, especially on the rate estimations and on the fitting procedure itself and Frederique Marion for many useful ideas and suggestions and for proof-reading the manuscript. This research was supported in part by PPARC grant PP/F001096/1.

\newpage
\appendix
\section{Norming the luminosity functions}
\label{app:norming}

This appendix shows the derivation of the proper norming for the luminosity models. In the following the substitution $x=L/L_0$ is used to calculate the norm $\Phi_0$.

\begin{enumerate}
 \item A single power law:

\begin{equation}
 \Phi_0 = \frac{(\alpha+1)L_0^\alpha}{(x^{\alpha+1}_\wedge-x^{\alpha+1}_\vee)}
\end{equation}

With this norming it is obvious that the value of the integral $\int_{L_*}^{L_\wedge} \Phi(L) dL$ actually does not depend on the value $L_0$, but only on the exponent $\alpha$.

\begin{eqnarray}
 \Phi(x) &=& \frac{x_\wedge^{\alpha+1}-x^{\alpha+1}}{x_\wedge^{\alpha+1}-x_\vee^{\alpha+1}}  \nonumber \\
&\equiv& \frac{1-x^{\alpha+1}}{1-x_\vee^{\alpha+1}}
\end{eqnarray}

\item The broken power law:

Integration of (\ref{eq:broken}) gives the following normalization constant:

\begin{equation}
 \Phi_0 = \left( \frac{ x_0^{\alpha+1} -x_\vee^{\alpha+1}}{\alpha+1} + \frac{x_\wedge^{\beta+1}-x_0^{\beta+1}}{\beta+1}  \right)^{-1}
\end{equation}

and so the complete expression of the integration of the luminosity function from a lower bound $x$ to $x_\wedge$ gives:

\begin{eqnarray}
 \int_{x}^{x_\wedge} \Phi(x) dx  &=& \left( \frac{ 1 -x_\vee^{\alpha+1}}{\alpha+1} + \frac{x_\wedge^{\beta+1}-1}{\beta+1}  \right)^{-1} \\ \nonumber
& \times & \left( \frac{ 1 -x^{\alpha+1}}{\alpha+1} + \frac{x_\wedge^{\beta+1}-1}{\beta+1}  \right) 
\end{eqnarray}

if the value $x$ lies within the interval $[x_\vee; x_0]\equiv[x_\vee; 1]$. If $x$ is larger than 1 the single power law can be used, scaled by a value $\kappa$ so that the value at the intersection point (which is $x\equiv 1$) falls together. Thus we require:

\begin{eqnarray}
 \Phi_0 \times \left( \frac{x_\wedge^{\beta+1}-1}{\beta+1}  \right) =
\kappa\frac{x_\wedge^{\beta+1}-1}{x_\wedge^{\beta+1} -x_\vee^{\beta+1} } 
\end{eqnarray}

from which follows:

\begin{equation}
 \kappa = \Phi_0 \frac{x_\wedge^{\beta+1}-x_\vee^{\beta+1}}{\beta+1} \;.
\end{equation}

\item The Schechter and the log-normal distribution

For the Schechter and the log-normal distribution the norming (i.e. the cumulative distribution) is computed numerically by numerical integration.

\end{enumerate}

\begin{figure}[h!]
\includegraphics[width=2.5in,angle=0]{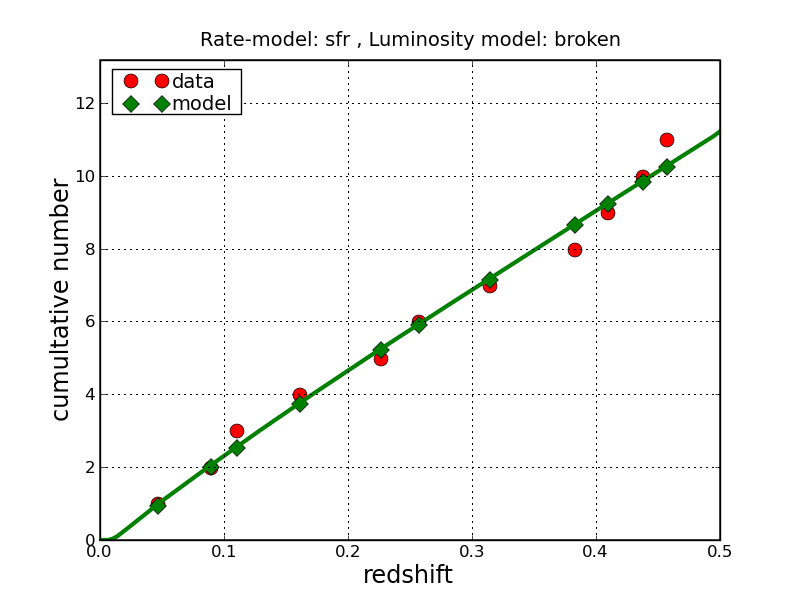}
\includegraphics[width=2.5in,angle=0]{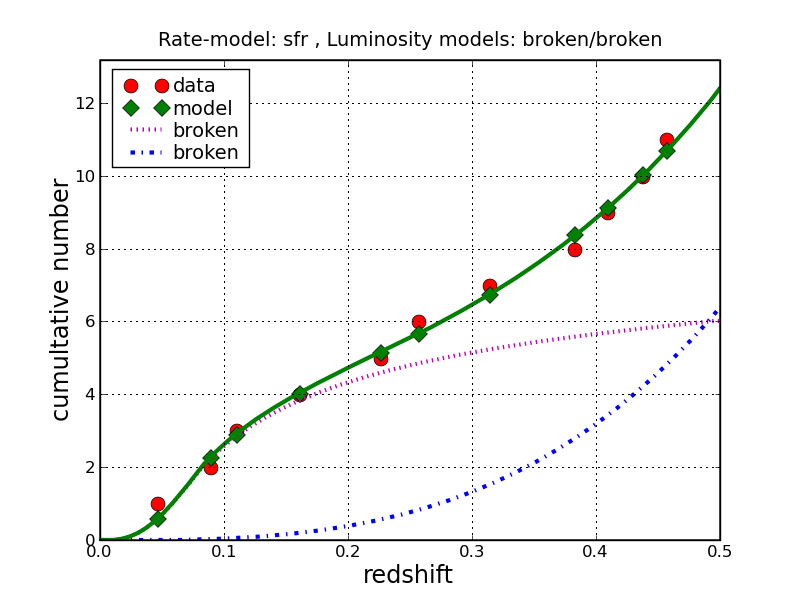}
\includegraphics[width=2.5in,angle=0]{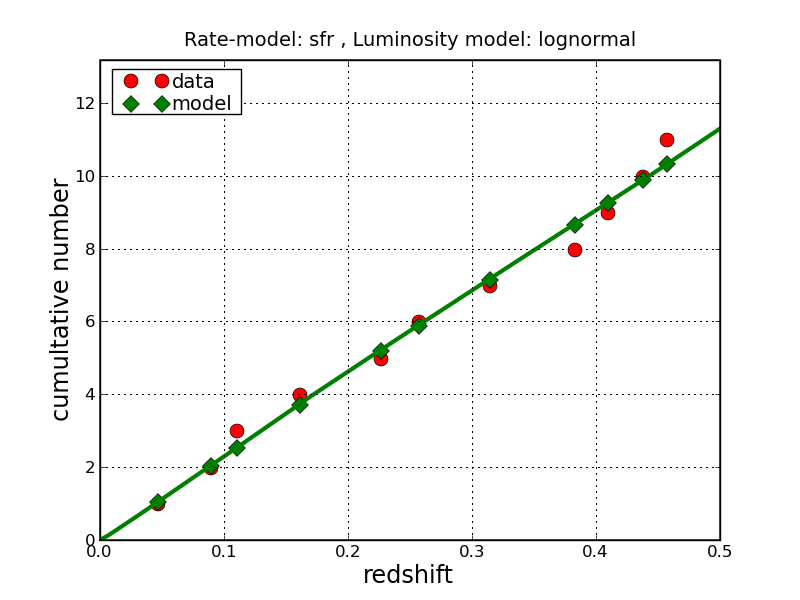}
\includegraphics[width=2.5in,angle=0]{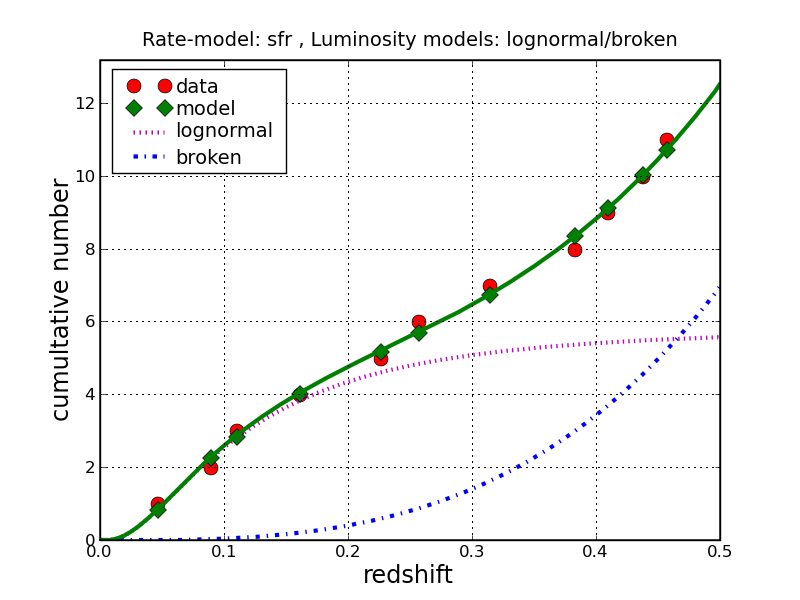}
\includegraphics[width=2.5in,angle=0]{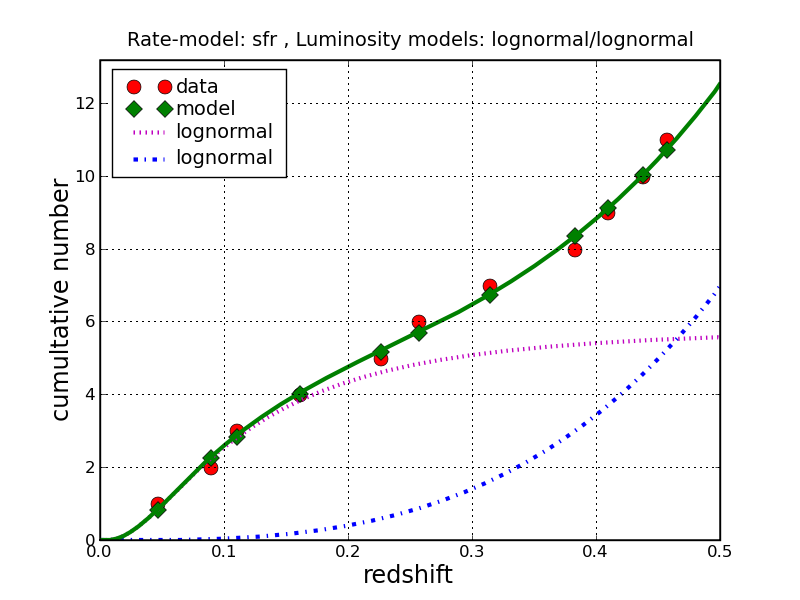}
\includegraphics[width=2.5in,angle=0]{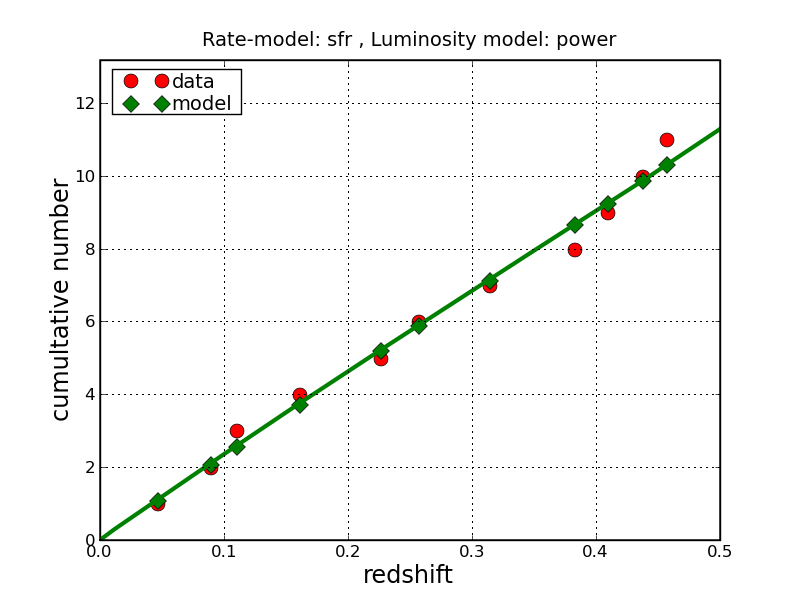}
\includegraphics[width=2.5in,angle=0]{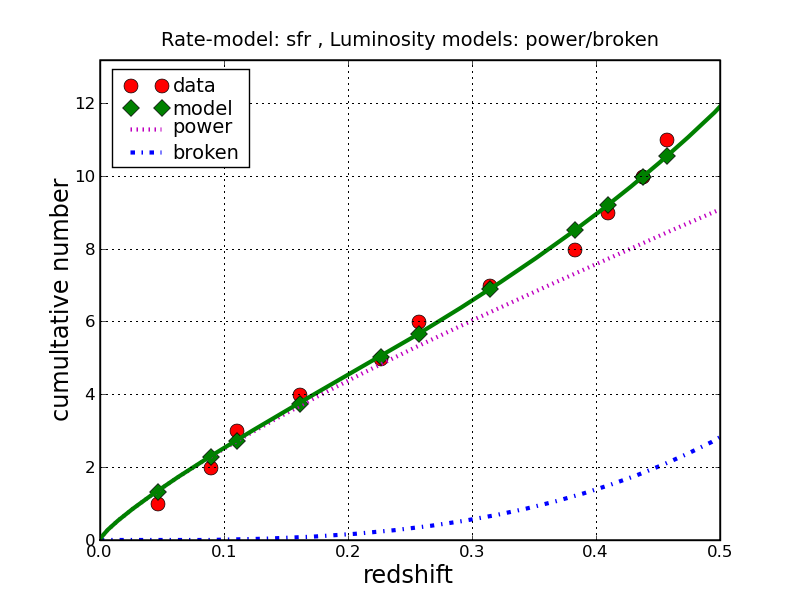}
\includegraphics[width=2.5in,angle=0]{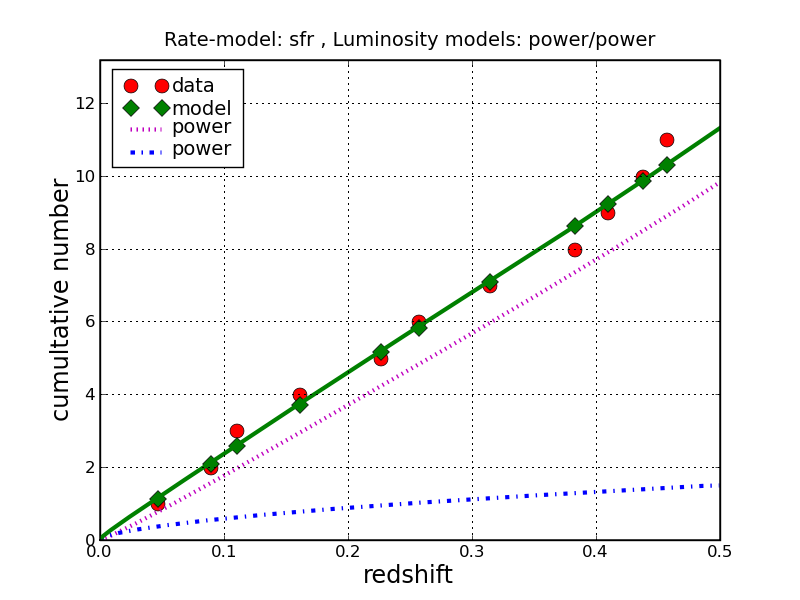}
\caption{Part I of the fits with the SF2 rate function. 
The x-axis shows the redshift and the y-axis the cumulative number of SGRB detected up to the given redshift.
The red dots represent the observed data while the green line shows the best-fit model. 
In case of a two-model fit the thinner red and blue lines represent the contributions from both models. 
}
\label{fig:imageFitsSFR1}
\end{figure}

\begin{figure}[h!]
\includegraphics[width=2.5in,angle=0]{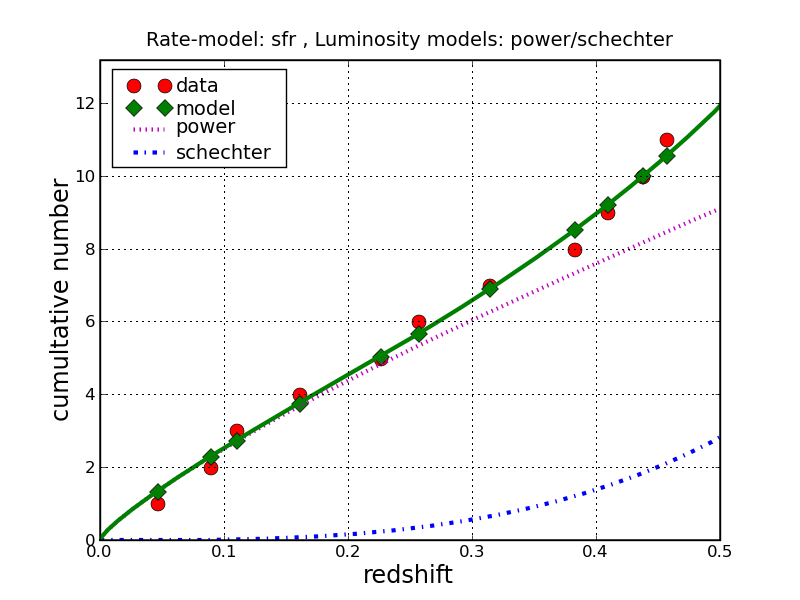}
\includegraphics[width=2.5in,angle=0]{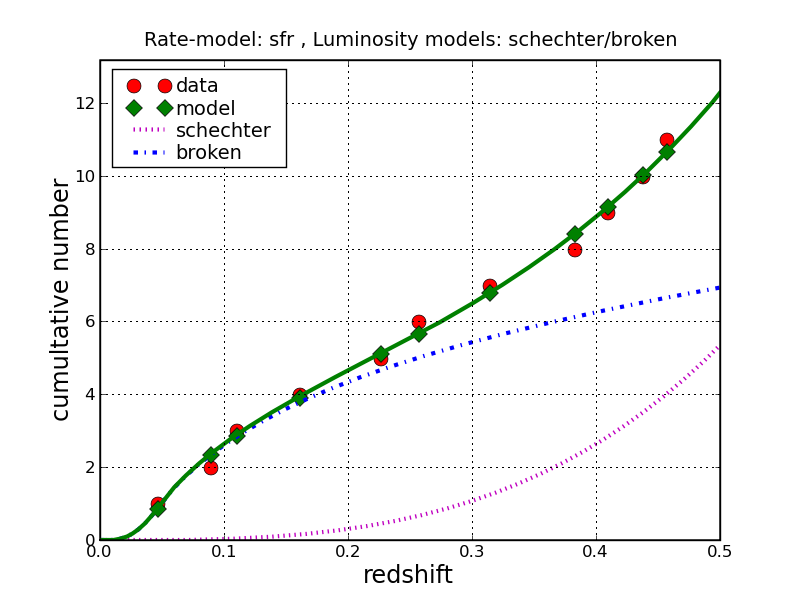}
\includegraphics[width=2.5in,angle=0]{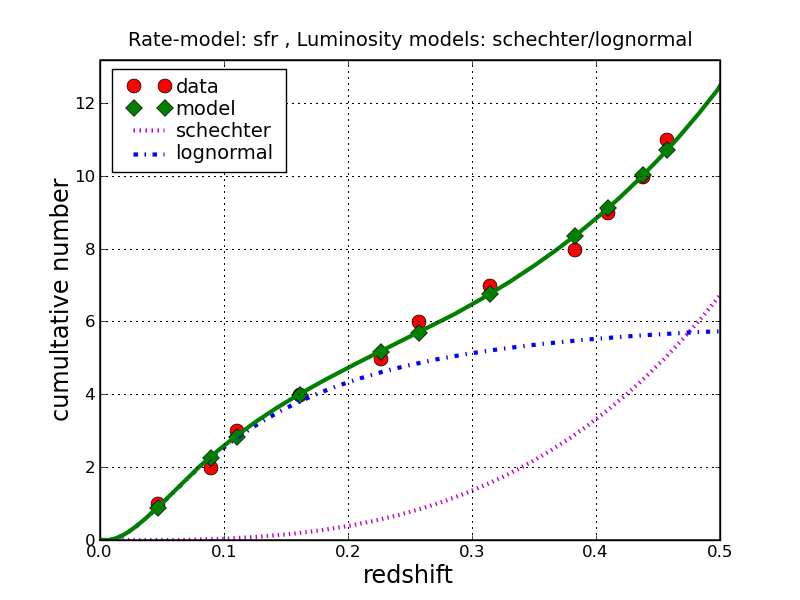}
\includegraphics[width=2.5in,angle=0]{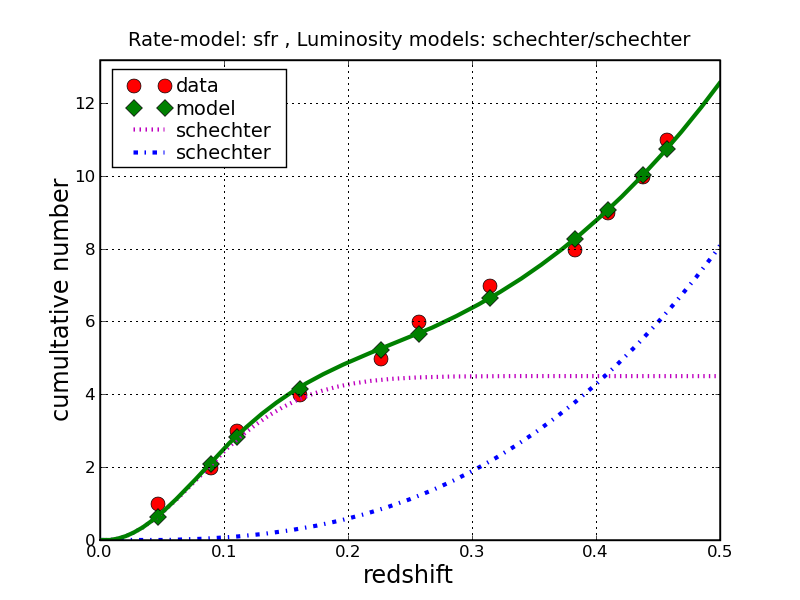}
\caption{Part II of the fits with the SF2 rate function. 
The x-axis shows the redshift and the y-axis the cumulative number of SGRB detected up to the given redshift.
The red dots represent the observed data while the green line shows the best-fit model. 
In case of a two-model fit the thinner red and blue lines represent the contributions from both models. 
}
\label{fig:imageFitsSFR2}
\end{figure}

\begin{figure}[h!]
\includegraphics[width=2.5in,angle=0]{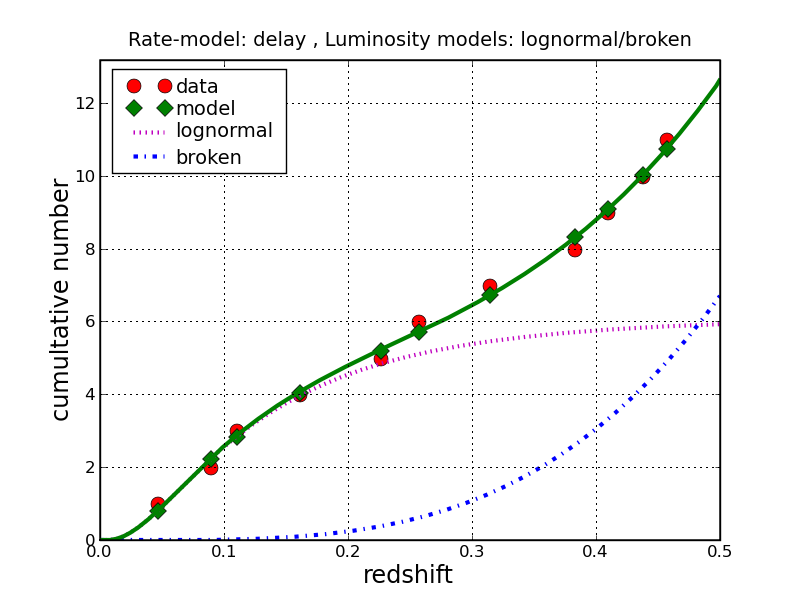}
\includegraphics[width=2.5in,angle=0]{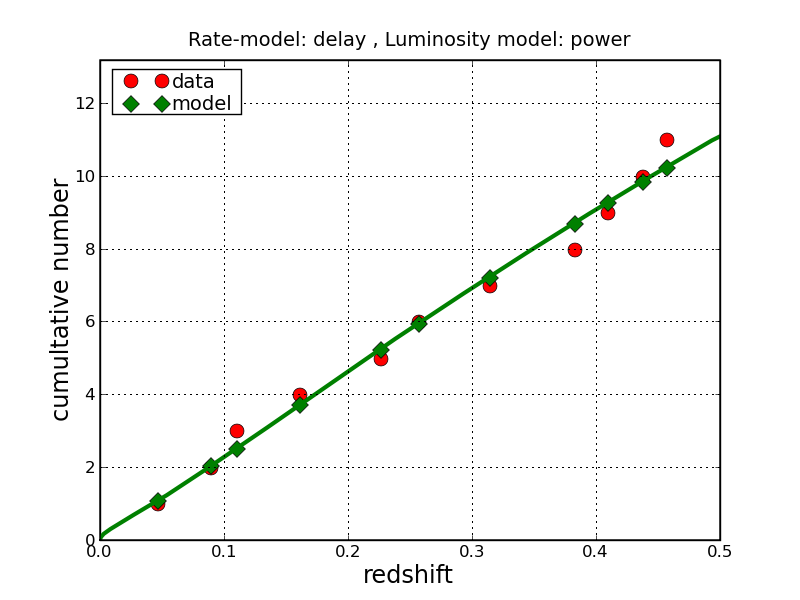}
\includegraphics[width=2.5in,angle=0]{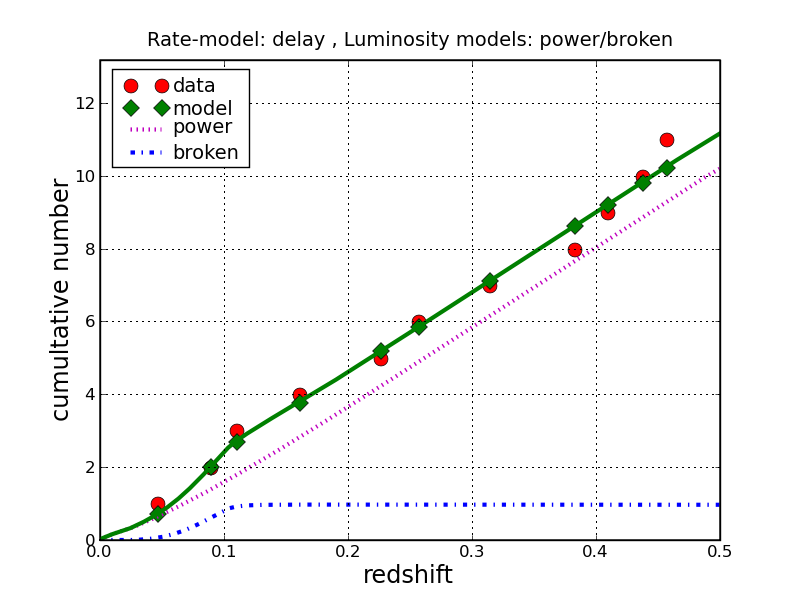}
\includegraphics[width=2.5in,angle=0]{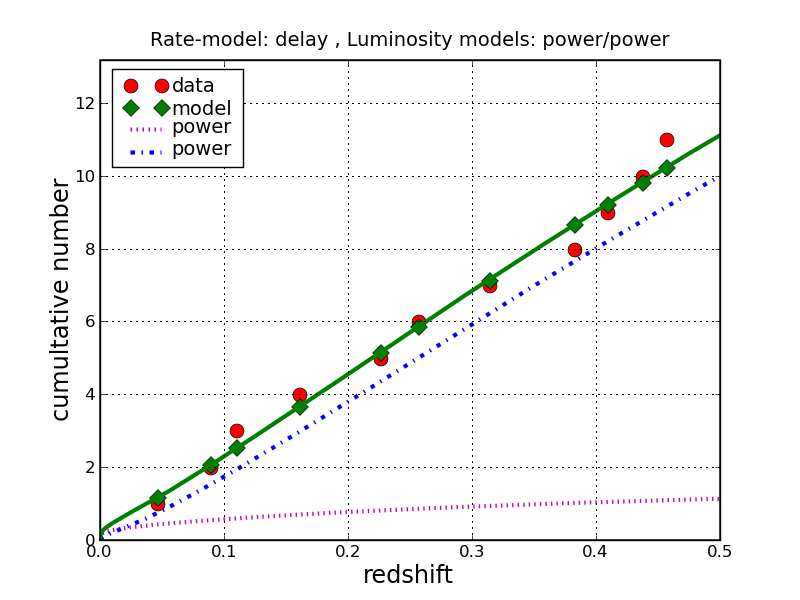}
\includegraphics[width=2.5in,angle=0]{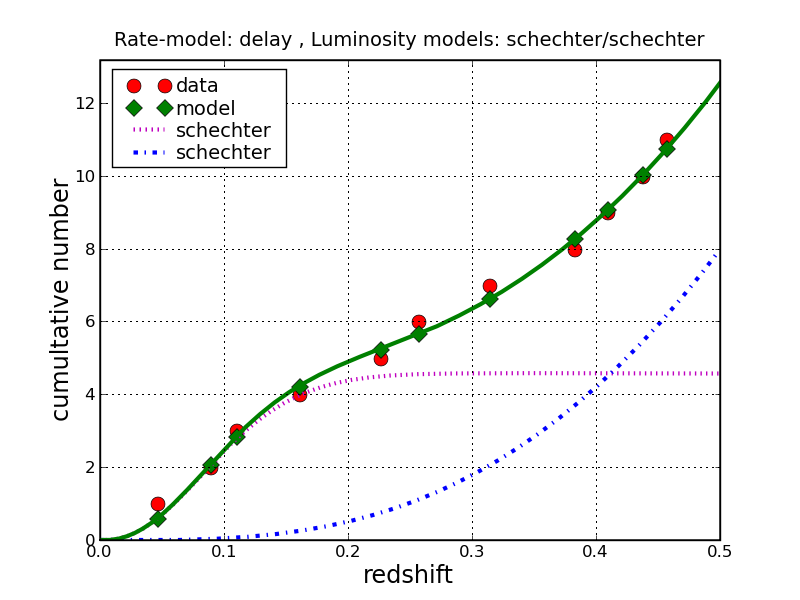}
\caption{Fits made with the delay rate function. 
The x-axis shows the redshift and the y-axis the cumulative number of SGRB detected up to the given redshift.
The red dots represent the observed data while the green line shows the best-fit model. 
In case of a two-model fit the thinner red and blue lines represent the contributions from both models. 
}
\label{fig:imageFitsDelay}
\end{figure}

\begin{figure}[h!]
\includegraphics[width=2.5in,angle=0]{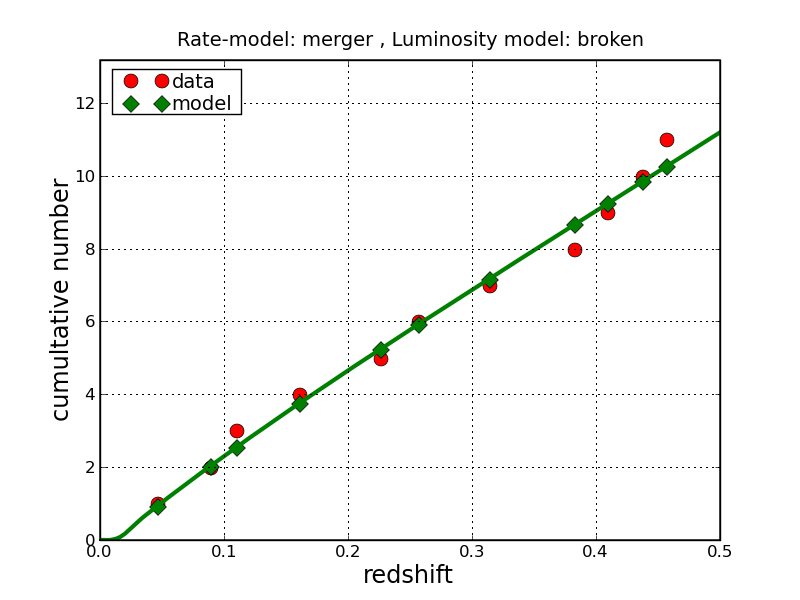}
\includegraphics[width=2.5in,angle=0]{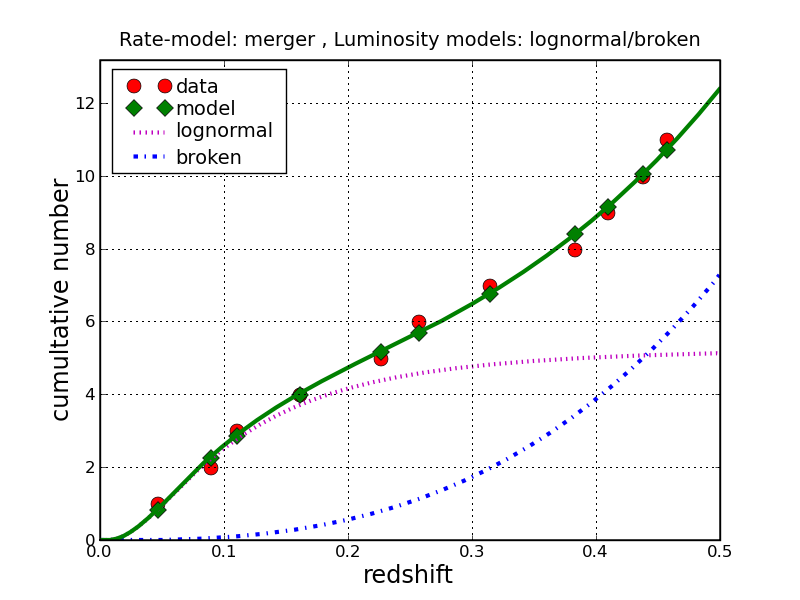}
\includegraphics[width=2.5in,angle=0]{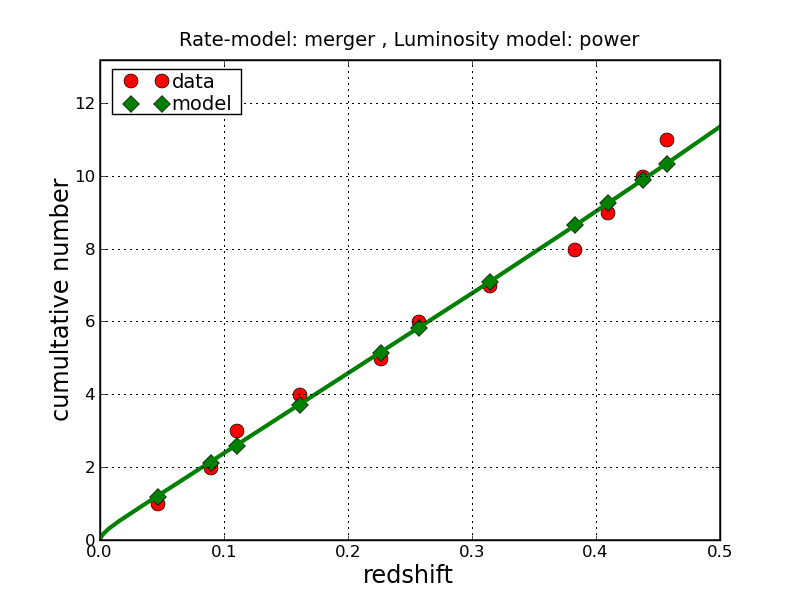}
\includegraphics[width=2.5in,angle=0]{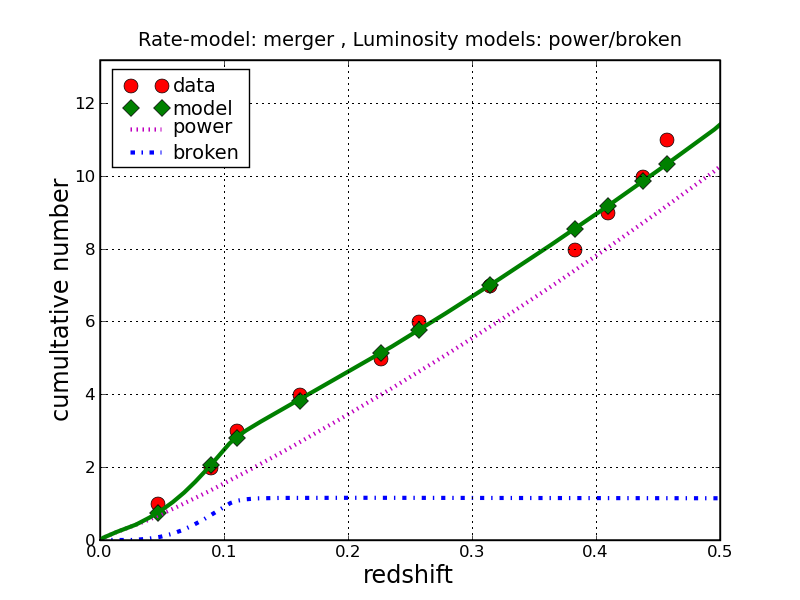}
\includegraphics[width=2.5in,angle=0]{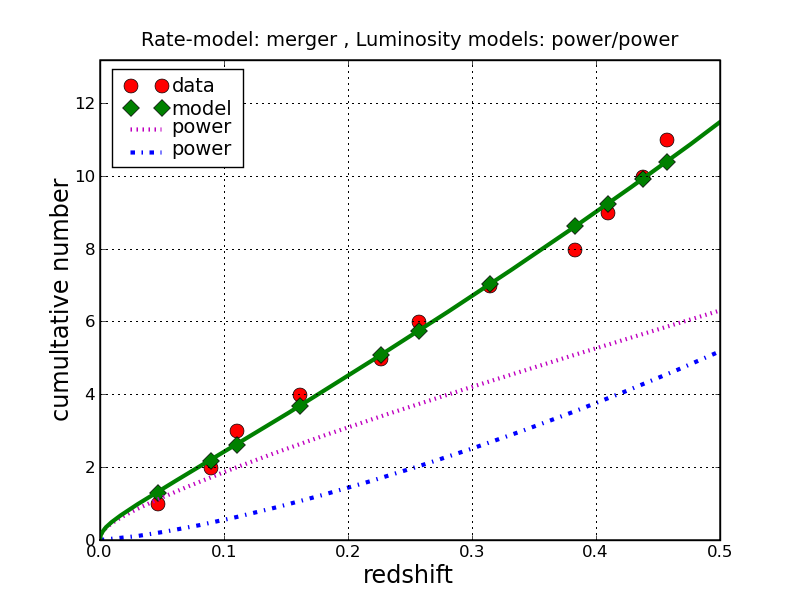}
\includegraphics[width=2.5in,angle=0]{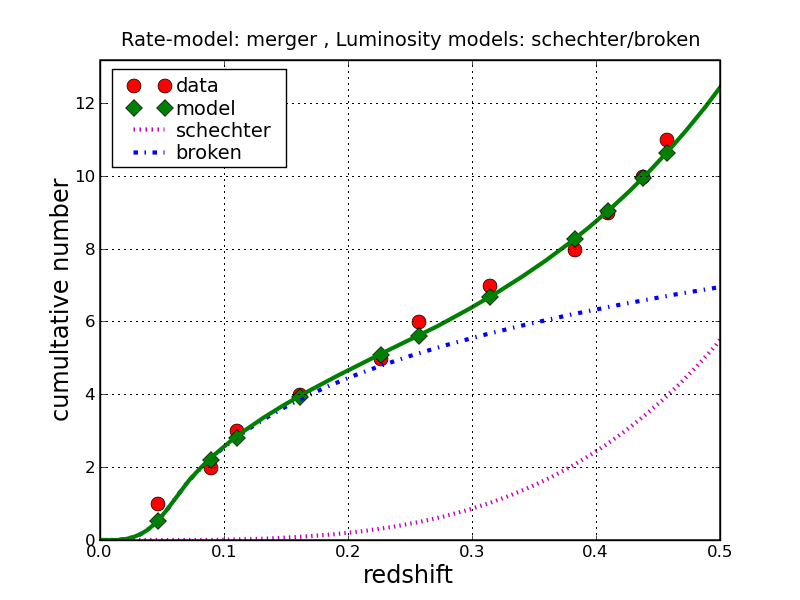}
\includegraphics[width=2.5in,angle=0]{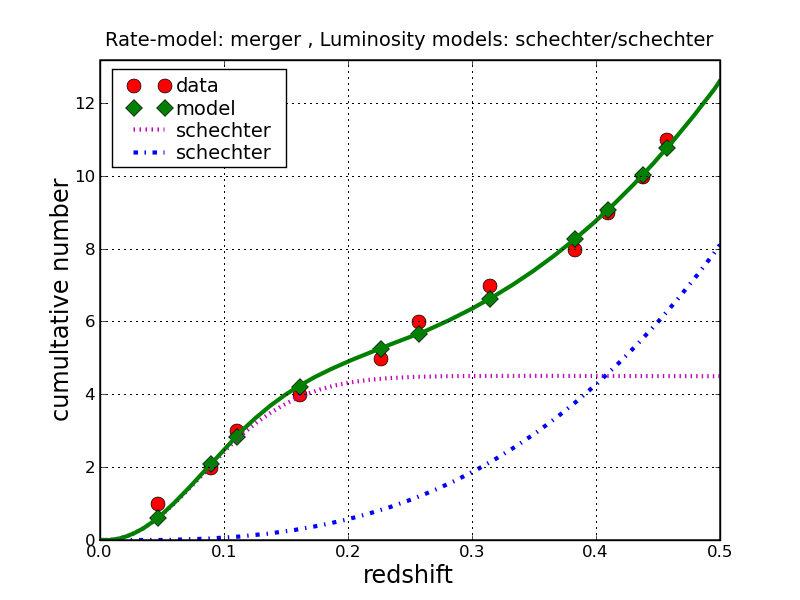}
\caption{Fits made with the merger rate function. 
The x-axis shows the redshift and the y-axis the cumulative number of SGRB detected up to the given redshift.
The red dots represent the observed data while the green line shows the best-fit model. 
In case of a two-model fit the thinner red and blue lines represent the contributions from both models. 
}
\label{fig:imageFitsMerger}
\end{figure}

\clearpage

\bibliography{/virgo/users/dietz/Documents/MyWork/Papers/Bibliography/bib,/virgo/users/dietz/Documents/MyWork/Papers/Bibliography/iulpapers}

\end{document}